\documentstyle[11pt,newpasp,twoside,psfig,epsf]{book}

\newcommand{\ie}{{\em i.e.},\ }
\newcommand{\eg}{{\em e.g.},\ }
\newcommand{\etal}{{\em et al.}\ }

\begin{document}

\pagenumbering{arabic}
\setcounter{page}{34}

%
%

\markboth{Brunner}{The New Paradigm}

\setcounter{section}{0}
\setcounter{figure}{0}
\setcounter{table}{0}
\setcounter{footnote}{0}

\title{The New Paradigm: Novel, Virtual Observatory Enabled Science}
\author{R.J. Brunner}
\affil{Department of Astronomy, California Institute of Technology, Pasadena, CA, 91125}

\begin{abstract}
A virtual observatory will not only enhance many current scientific
investigations, but it will also enable entirely new scientific
explorations due to both the federation of vast amounts of
multiwavelength data and the new archival services which will, as a
necessity, be developed. The detailing of specific science use cases
is important in order to properly facilitate the development of the
necessary infrastructure of a virtual observatory. The understanding
of high velocity clouds is presented as an example science use case,
demonstrating the future synergy between the data (either catalog or
images), and the desired analysis in the new paradigm of a virtual
observatory.

\end{abstract}

\section{Introduction}

A diverse and exciting array of scientific possibilities, whose
exploration are enhanced by the existence of a Virtual Observatory,
are detailed elsewhere in this volume. Certain lines of scientific
inquiry, however, are not just enhanced by a virtual observatory, but
are actually enabled by it. For example, a panchromatic study of
active galactic nuclei (see, \eg Boroson, these proceedings), studies
of the low surface brightness universe (see, \eg Schombert, these
proceedings), a study of Galactic structure (see, \eg Kent, these
proceedings), or a panchromatic study of galaxy clusters, are all
extremely interesting projects that are facilitated by a virtual
observatory.

In this article, I will discuss some specific technical challenges
which must be overcome in order to fully enable this new type of
scientific inquiry. This is not as difficult as it may first appear,
as many of these challenges are already being tackled, as is evidenced
by the prototype services which are currently available at many of the
leading data centers. In order to truly make revolutionary, and not
merely evolutionary, leaps forward in our ability to answer the
important scientific questions of our time, we need to ``think outside
the box'', not just in the design and implementation of a virtual
observatory, but in the actual scientific methodology we wish to
employ (see, \eg Figure 1, which demonstrates this concept by
combining large image viewing with the ability to selectively mark
objects in the image based on their statistical properties).

\begin{figure}[!htb]
\plotfiddle{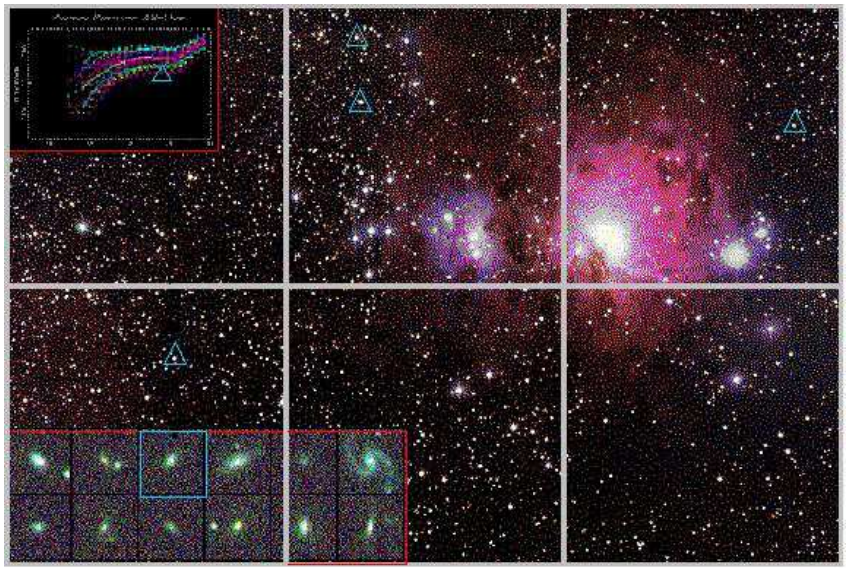}{3.0in}{0}{55}{55}{-162}{0}
\caption{A prototype of the visualization services which would empower 
scientists to not only tackle current scientific challenges, but also
to actually aid in the exploration of the, as yet unknown, challenges
of the future. Note the intelligent combination of image and catalog
visualizations to aid the scientist in exploring parameter space. Figure
courtesy of Joe Jacob, JPL.}
\end{figure}

\section{Technological Challenges}

While many of the technical challenges are rather self-evident upon a
cursory examination, such as the federation of existing archival
centers, other challenges are considerably more difficult to
elucidate. This effect is primarily a result of the difficulty in
designing scientific programs for the, as yet unavailable, virtual
observatory. This is exactly the time where ``thinking outside the
box'' applies, as one needs to ask not ``{\em what can I do right now?}'',
but ``{\em what would I like to be able to do?}''. 

The first step in this process is to consider, in its entirety, all of
the data which might be available for ingestion into a virtual
observatory. This includes the obligatory data catalogs, which are the
most often used derivative of survey programs, and perhaps more
importantly, the original imaging data and any associated metadata
(that is, data which describes the data). Similar extensions likewise
apply to other types of astronomical data, including spectral and
temporal.

After taking this revolutionary leap, we can now consider querying not
just catalogs, but also the data from which the catalogs were
extracted. This would allow for new techniques to be applied, which
might, for example, perform source extraction using multiple
wavelength images simultaneously (\eg $\chi^2$ detection, Szalay \etal
1999), or perhaps to extract flux limits for objects detected in other
wavelengths, or, finally, to extract matched parameters (\eg matched
aperture photometry).

\begin{figure}[!htb]
\plotfiddle{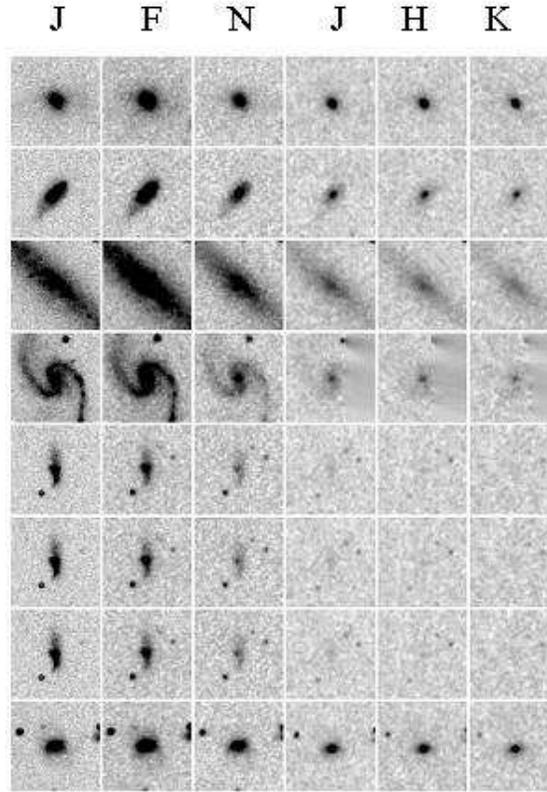}{3.95in}{0}{70}{70}{-108}{0}
\caption{The multiwavelength nature of nearby galaxies, constructed 
from the optical data of the DPOSS survey (J, F, and N) and the
near-infrared data of the 2MASS survey (J, H, and K). Figure
courtesy of Steve Odewahn, Arizona State University.}
\end{figure}

This is demonstrated in Figure 2, where the multiwavelength nature of
nearby galaxies is explored, from the optical, extracted from the
DPOSS survey (Djorgovski \etal 1998), to the near-infrared, extracted
form the 2MASS survey (Skrutskie \etal 1997). As this example
demonstrates, multiwavelength image processing is a pressing need,
since objects bright in one wavelength are often much fainter, if even
detected at all, at other wavelengths.

\begin{figure}[!htb]
\plottwo{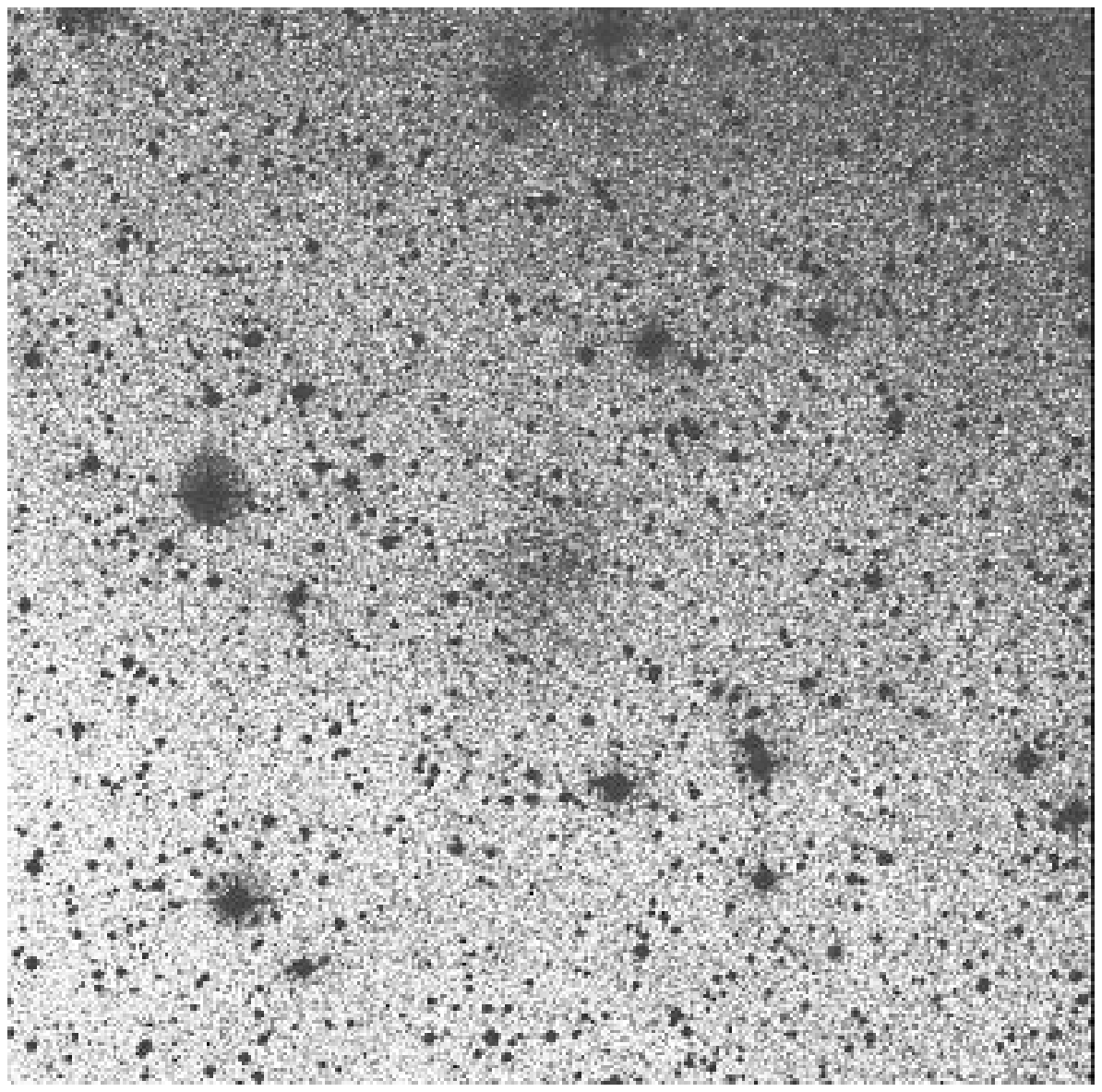}{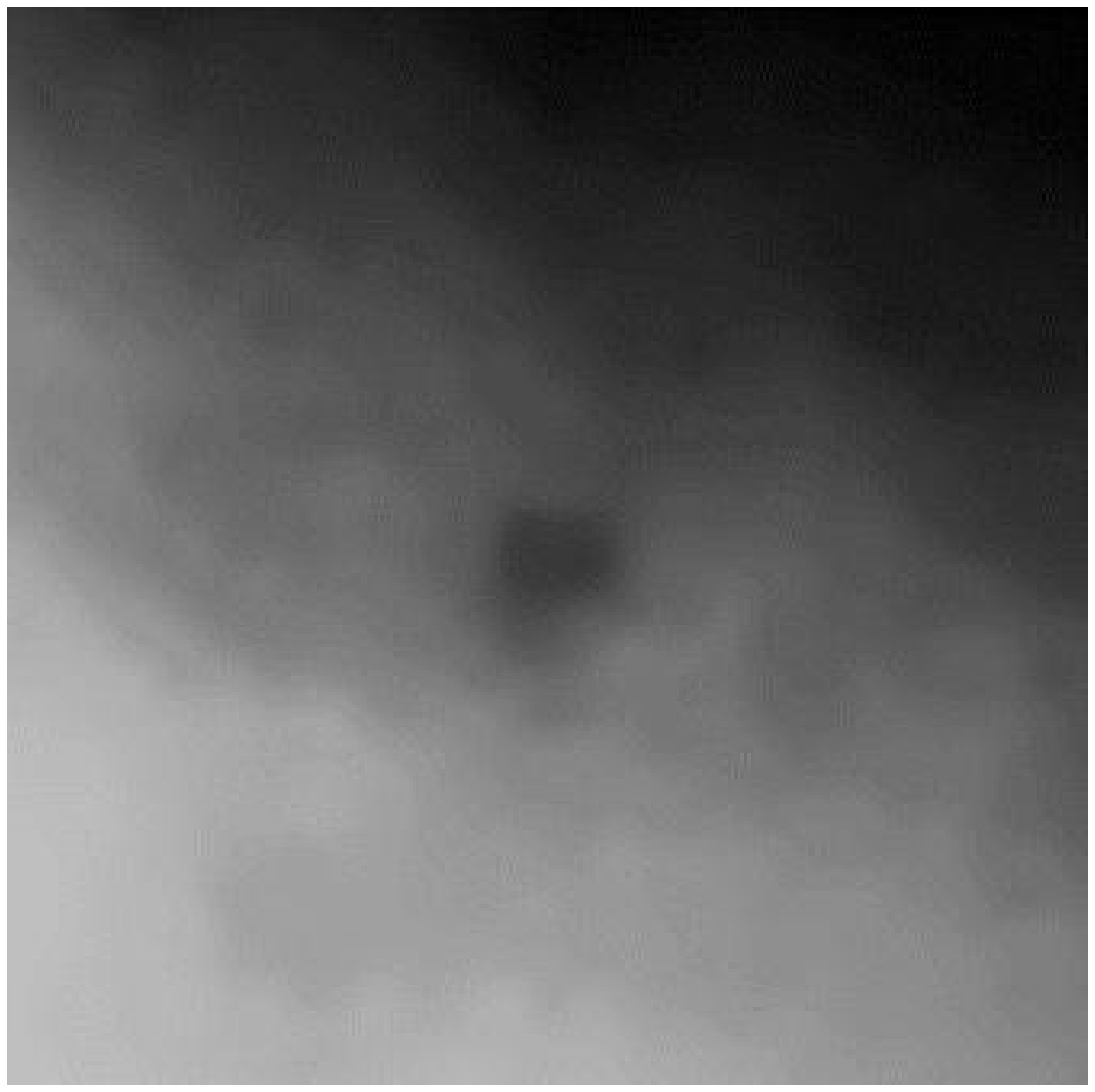}
\caption{A demonstration of the detection process which must be 
used to detect the low surface brightness sources missed by most
survey pipelines. On the left is the original DPOSS plate image which
contains the dwarf spheroidal galaxy Andromeda V (Armandroff \etal
1998), which is visible with a large image stretch.  On the right is a
reprocessed version of the original image which is designed to
emphasize subtle background variations which can be caused by low
surface brightness sources (see, \eg Brunner \etal 2001).}
\end{figure}

Another example of the need for image reprocessing is shown in Figure
3, where the detection of a nearby, low surface brightness dwarf
spheroidal galaxy is demonstrated. The vast majority of survey
pipelines are designed to detect the dominant source population,
namely high surface brightness point-type sources. As a result, an
implicit surface-brightness selection effect exists in nearly all
catalogs (see, \eg Schombert, these proceedings for a more detailed
account). In the future, one would ideally like to be able to
reprocess survey data in an effort to find objects at varying spatial
scales and surface brightnesses.
 
\section{Science Use Case: Understanding High Velocity Clouds}

As a demonstration of how a virtual observatory can enable new
science, consider the specific science use case of understanding high
velocity clouds (HVCs). HVCs are defined as systems consisting of
neutral Hydrogen which have velocities that are incompatible with
simple models of Galactic rotation (Wakker and van Woerden 1997).
Their origin, however, remains uncertain, with various arguments being
made in support of a wide range of hypothesis, including that they are
Galactic constituents, that they are the remnants of galaxy interactions,
or that they are fragments from the hierarchical formation of our
local group of galaxies.

In an effort to truly understand these systems, we also would like to
understand their composition. Although these systems are, by
definition, found in neutral Hydrogen surveys, we can perform either
follow-up observations at other wavelengths, or else correlate the HI
data with existing surveys at other wavelengths in order to learn more
about them (see, \eg Figure 4 for a demonstration of multiwavelength
image correlation for a known HVC). This process can often require the
construction of large image mosaics involving multiple POSS-II
photographic plates in order to map structures that span several tens
of square degrees. This service should clearly be one of the principal
design requirements for a virtual observatory.

The most powerful method for understanding the composition of HVCs,
however, is to study their absorption effects on the spectra of
background sources, most notably quasars. In order to find suitable
targets, we need to be able to dynamically correlate the HVC images
with published quasar catalogs in order to determine the optimal
line-of-sights for quantifying the composition of the intervening
HVCs with follow-up spectral observations.

\begin{figure}[!htb]
\plottwo{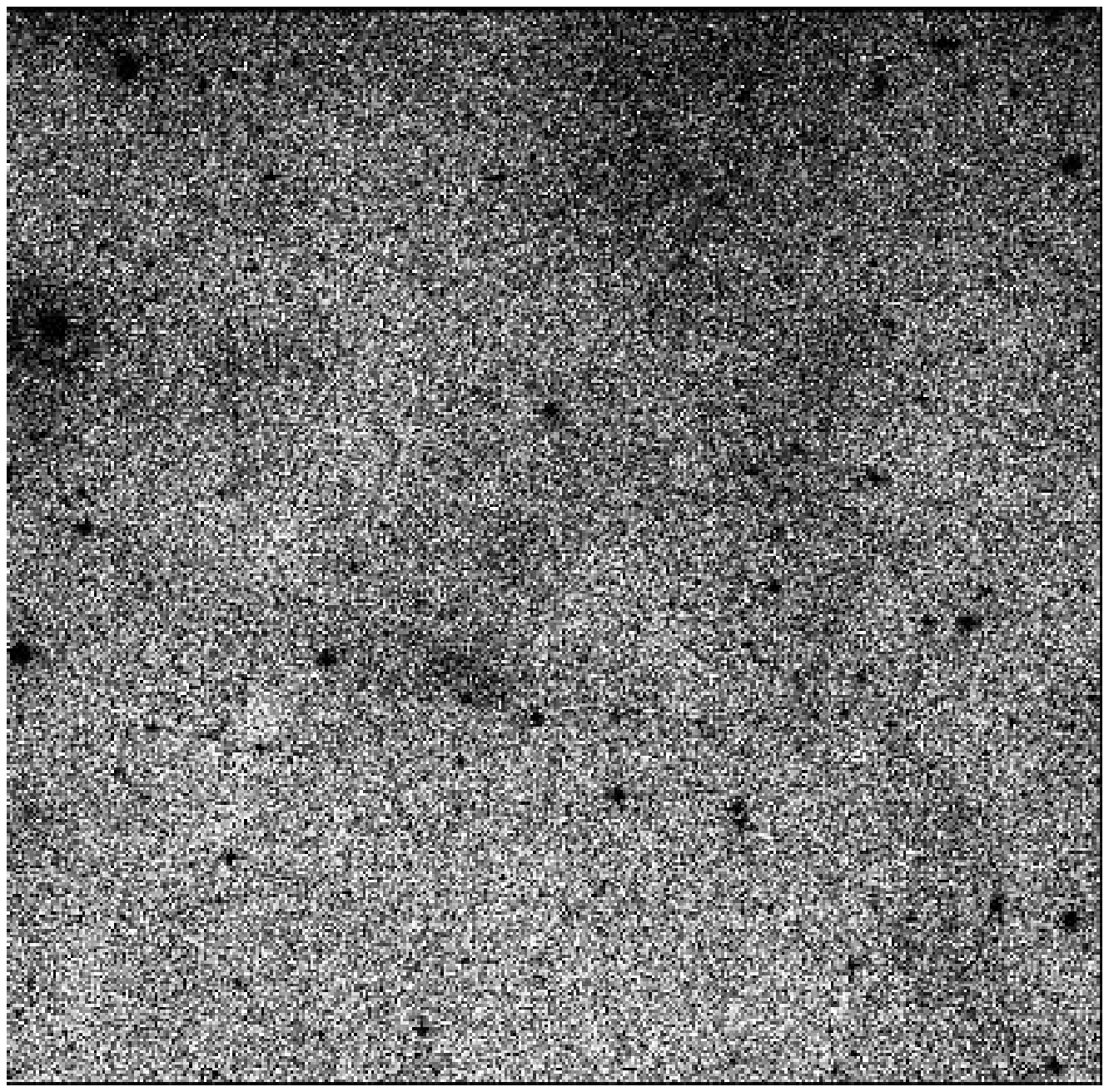}{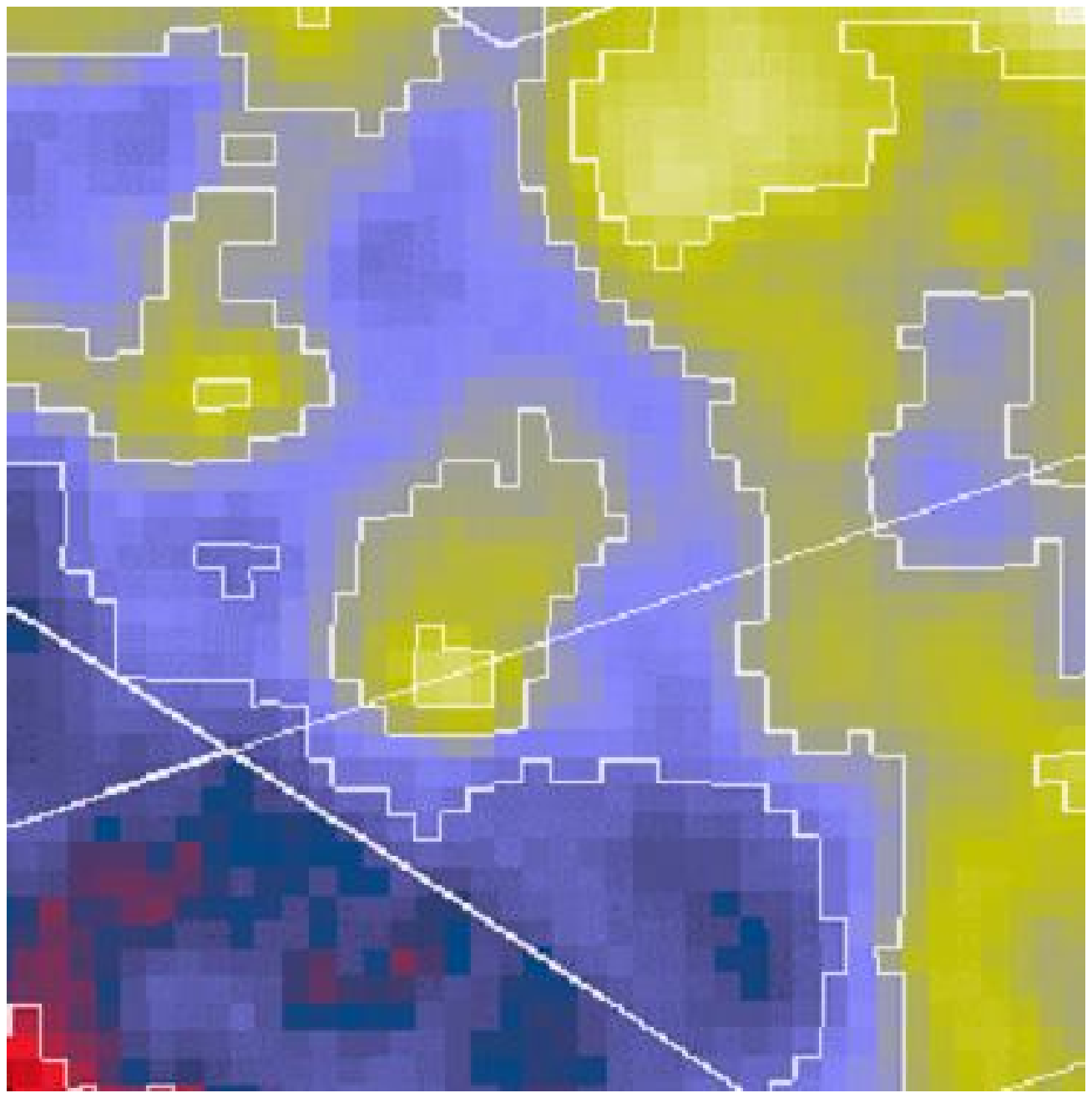}
\caption{A demonstration of the image correlation process for a known 
high velocity cloud complex. The figure on the left is a DPOSS F
(approximately R) plate image, selected since it would contain any
$H_{\alpha}$ emission. The HVC is detected by the low surface
brightness emission near the center of the image. The figure on the
right is the correlated IRAS $100 \micron$ image, which demonstrates a
remarkable correlation with the diffuse emission seen in the figure on
the right (see Brunner \etal 2001).}
\end{figure}

Finally, we also would like to understand the evolution of these
systems, which has obvious implications for understanding their
origin. This can optimally be done by comparing the predictions of
theoretical models to our correlated multiwavelength
observations. This implies a need for a virtual observatory to allow
seamless access to not only astronomical data but also the results of
dynamical analysis, either through persisted calculations or a
real-time process.

\section{The New Paradigm}

To accomplish these ambitious scientific goals, we need powerful
tools, which should be implemented as part of a virtual
observatory. First, we need the ability to process and visualize large
amounts of imaging data. This should be done in both a manner which is
suitable for public consumption (\ie the virtualsky.org project) and
also a manner which preserves scientific calibrations. These services
will also need to provide coordinate transformations, overlays and
arbitrary re-pixelizations. Ideally, these operations occur as part of a
service which can also accept user-defined functions to further
process the data, minimizing the size of the data stream which must be
established with the end-user.

Next, we need the ability to federate an arbitrary collection of
catalogs, selected from geographically diverse archives, a prime
computational grid application. To completely enable the discovery
process, we also need intelligent display mechanisms to explore the
high-dimensionality spaces which will result from this federation
process.  We also will need to allow the user to post-process these
federations using user-defined tools or functions (\eg statistical
analysis) as well as combine these processes with image operations and
visualizations.

Finally, a complete census and subsequent description of science use cases (\eg
the previous section, see also, Boroson, these proceedings),
inevitably leads one to the formulation of a new paradigm for doing
astronomy with a virtual observatory. In the future, anyone, anywhere,
will be able to do cutting edge science, as researchers will only be
limited by their creativity and energy, not their access to
restricted observations or telescopes. Not only will this revolutionize
the scientific output of our community, but it will also have an
important effect on the sociology of our field as well, since students
will need to be trained in these new tools and techniques.

\acknowledgements

This work was made possible in part through the NPACI sponsored
Digital Sky project and a generous equipment grant from SUN
Microsystems. RJB would like to acknowledge the generous support of
the Fullam Award for facilitating this project. Access to the DPOSS
image data stored on the HPSS, located at the California Institute of
Technology, was provided by the Center for Advanced Computing
Research. The processing of the DPOSS data was supported by a generous
gift from the Norris foundation, and by other private donors.


\begin{references}

\reference Armandroff, T.E., Davies, J.E., \& Jacoby, G.H. 1998, \aj, 116, 2287

\reference Brunner, R.J., Roth, N., Hokada, J., Gal, R., Mahabal, A.A., Odewahn, S.C.,
\& Djorgovski, S.G. 2001, \aj, in preparation

\reference Djorgovski, S., de Carvalho, R.R., Gal, R.R., Pahre, M.A., 
Scaramella, R., and Longo, G. 1998, In B. McLean, editors, {\em The
Proceedings of the $179^{th}$ IAU on New Horizons from
Multi-Wavelength Sky Surveys}, IAU Symposium No. 179, 424

\reference Skrutskie, M.F., \etal 1997, In F. Garzon \etal, editors, 
{\em The Impact of Large Scale Near-IR Sky Surveys}, Kluwer, 25

\reference Szalay, A.S., Connolly, A.J., \& Szokoly, G.P. 1999, \aj, 117, 68 

\reference Wakker, B.P., \& van Woerden, H. 1997, \araa, 35, 217

\end{references}
\end{document}